\documentclass[graybox]{svmult}

\usepackage{mathptmx}       
\usepackage{helvet}         
\usepackage{courier}        
\usepackage{type1cm}        
\usepackage{makeidx}         
\usepackage{graphicx}        
\usepackage{multicol}        
\usepackage[bottom]{footmisc}

\newcommand {\apgt} {\ {\raise-.5ex\hbox{$\buildrel>\over\sim$}}\ }
\newcommand {\aplt} {\ {\raise-.5ex\hbox{$\buildrel<\over\sim$}}\ }

\makeindex             

\begin{document}

\title*{The scientific potential of space-based gravitational wave detectors}
\author{Jonathan R Gair}
\institute{Jonathan R Gair \at Institute of Astronomy, University of Cambridge, Madingley Road, Cambridge, CB3 0HA, UK \email{jgair@ast.cam.ac.uk}}
%
%
\maketitle

\abstract{The millihertz gravitational wave band can only be accessed with a space-based interferometer, but it is one of the richest in potential sources. Observations in this band have amazing scientific potential. The mergers between massive black holes with mass in the range $10^4-10^7M_\odot$, which are expected to occur following the mergers of their host galaxies, produce strong millihertz gravitational radiation. Observations of these systems will trace the hierarchical assembly of structure in the Universe in a mass range that is very difficult to probe electromagnetically. Stellar mass compact objects falling into such black holes in the centres of galaxies generate detectable gravitational radiation for several years prior to the final plunge and merger with the central black hole. Measurements of these systems offer an unprecedented opportunity to probe the predictions of general relativity in the strong-field and dynamical regime. Millihertz gravitational waves are also generated by millions of ultra-compact binaries in the Milky Way, providing a new way to probe galactic stellar populations. ESA has recognised this great scientific potential by selecting {\it The Gravitational Universe} as its theme for the L3 large satellite mission, scheduled for launch in $\sim 2034$. In this article we will review the likely sources for millihertz gravitational wave detectors and describe the wide applications that observations of these sources could have for astrophysics, cosmology and fundamental physics.}

\section{Introduction}
\label{sec:intro}
Gravitational waves (GWs) are expected to be generated in all frequency bands by a wide range of different sources. Over the next two decades several different frequency bands will be opened up observationally by a variety of different instruments. Extremely-low frequency gravitational waves ($\sim 10^{-16}$Hz) generated by quantum fluctuations during inflation imprint polarisation on the cosmic microwave background (CMB) which could be detected by ongoing experiments. In March 2014 the BICEP2 collaboration announced a detection of B-mode polarisation based on measurements of one part of the sky which were consistent with being produced by primordial gravitational radiation~\cite{bicep2}. These results are not yet confirmed and could be explained by foreground contamination, but they suggest that the signature of inflationary gravitational waves in the CMB could soon be detected. In the nanohertz frequency band, gravitational waves can be detected by the accurate timing of millisecond pulsars. These objects rotate very stably and so the time of arrival of pulses can be predicted very accurately. A gravitational wave propagating between the observer and the pulsar will cause the pulse to arrive earlier or later than expected and so gravitational waves can be detected by looking for periodic fluctuations in the pulse arrival times that are correlated between different pulsars on the sky. Three major pulsar timing array (PTA) efforts are currently underway --- the European PTA, NANOGrav, and the Parkes PTA --- and agreements exist to combine the data to create an international PTA~\cite{ipta}. GWs in the $10-1000$Hz band can be detected using kilometre-scale ground-based laser interferometers. A network of such detectors has been constructed and taken data over the last two decades (LIGO, Virgo, GEO, Kagra). No detections have yet been made, but scientifically interesting upper limits have been set and these detectors are currently undergoing upgrades to advanced configurations, with a factor of $\sim 10$ improvement in sensitivity, which will gradually start to take data from 2015~\cite{ligo}. These instruments could detect gravitational waves from coalescing compact binaries, from deformed rotating neutron stars, from transient events such as supernova or cosmic string cusps and perhaps from a stochastic background of radiation generated in the early Universe.

A frequency band missing from this list is the millihertz band from $\sim10^{-4}$--$10^{-1}$Hz. This frequency range is inaccessible from the ground due to the large seismic noise background that exists below $\sim 10$Hz and therefore to open it up we must put a detector in space. Proposals for a space-based interferometer have been discussed for over twenty years and it will be another twenty years before such an instrument is realised. However, this prospect has become more certain since ESA selected {\it The Gravitational Universe}~\cite{tgu} as the science theme to be addressed by the L3 large science mission due to be launched in $2034$. In this article we will describe the scientific potential of such an instrument. This article is organised as follows. In the remaining sections of this introduction we will describe millihertz gravitational wave detectors (Section~\ref{sec:detectors}) and potential sources of millihertz GW radiation (Section~\ref{sec:sources}). The remainder of the article will describe the scientific applications of millihertz GW observations to astrophysics (Section~\ref{sec:astro}), cosmology (Section~\ref{sec:cos}) and fundamental physics (Section~\ref{sec:fundphys}). We finish with a summary in Section~\ref{sec:discuss}.

\subsection{Detectors}
\label{sec:detectors}
\begin{figure}[b]
\sidecaption
\includegraphics[width=0.99\textwidth]{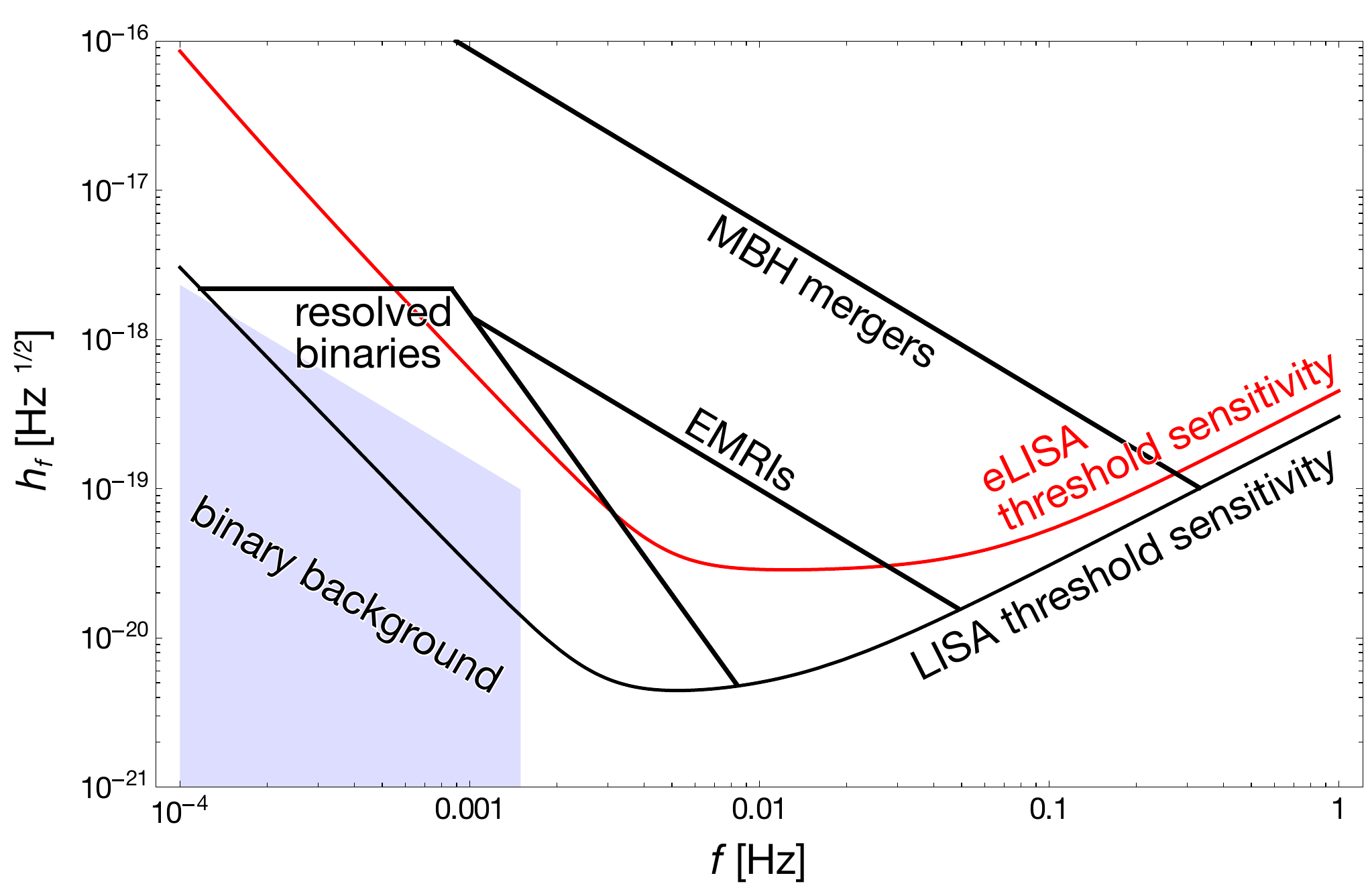}
%
%
\caption{Sensitivity curves for LISA and eLISA, with the approximate frequency range and amplitude of expected sources indicated. Reproduced from~\cite{TestGRLRR}.}
\label{fig:senscurve}       
\end{figure}

The canonical concept for a space-based millihertz gravitational wave detector is the Laser Interferometer Space Antenna (LISA)~\cite{lisa}. This design calls for a constellation of three spacecraft, arranged at the corners of an equilateral triangle, 5 million km apart in a heliocentric Earth-trailing orbit, $20^\circ$ behind the Earth. Two laser beams would pass along each arm of the triangle, one in each direction, and these would be used to make one-way measurements of the distance between the satellites from which the presence of gravitational waves can be inferred. LISA was developed as a joint NASA-ESA project until $2011$ when NASA withdrew from the mission due to funding constraints. A rescoped concept, NGO, was proposed as a candidate for the ESA-only L1 mission opportunity~\cite{eLISA,eLISAGWN}, but was not selected. NGO differed from LISA in the arm length --- 1 million km; in the number of arms --- only two of the sides of the constellation would have laser links instead of three; in the orbit --- $9^\circ$ behind the Earth and drifting away; in the mission duration --- $2$ years instead of $5$; and in specifics such as telescope diameter (20cm instead of 40cm) and acceleration noise requirements ($3$fm s$^{-2}$ Hz$^{-\frac{1}{2}}$, about a factor of $5$ less stringent than the LISA baseline). The primary difference between a two-arm and three-arm configuration is that the three-arm configuration allows the construction of two independent data streams, while only one can be constructed in the two-arm case, which has important consequences for parameter estimation. These differences allowed NGO to fit within the L1 mission budget cap, but lead to significant changes in sensitivity (see Figure~\ref{fig:senscurve}). {\it The Gravitational Universe}~\cite{tgu} science theme selected by ESA for the L3 mission used essentially the same NGO concept to illustrate the scientific potential. This mission is now referred to as eLISA (evolved-LISA) to reflect the fact that the design has not yet been fixed. We will refer to both LISA and eLISA in this article, but in all cases scientific results quoted for eLISA have been calculated using the NGO concept.

\subsection{Sources}
\label{sec:sources}
We describe here the four major types of millihertz gravitational wave source.
\subsubsection{Massive black hole mergers}
\label{srcsec:mbh}
Massive black holes (MBHs) are ubiquitous in the centres of galaxies~\cite{croton}. When galaxies merge it is expected that their black holes will sink to the centre of the merged galaxy through dynamical friction and eventually merge via GW emission. If the MBH has mass in the range $10^4$--$10^7M\odot$ the emitted radiation will be in the millihertz band. eLISA would be able to observe these merger with very high signal to noise ratio (S/N) anywhere in the observable Universe, for instance a merger of two $10^5M_\odot$ black holes would be observable with S/N or $50$ at a redshift of $z=20$~\cite{eLISA,eLISAGWN}. LISA would have been expected to observe a few tens of MBH merger events per year in `heavy seed' models and about twice as many events under `light seed' models~\cite{PEtask}. The expected event rate for eLISA is comparable for `heavy seed' models and a factor of $\sim 1.5$ lower for `light seed' models, as eLISA will not be able to observe the mergers between lighter black holes at higher redshift that make up a significant fraction of the LISA events in that case~\cite{eLISA,eLISAGWN}. The ability of a space-based GW detector to distinguish these and other models will be discussed in detail in Section~\ref{sec:astro}.

\subsubsection{Extreme-mass-ratio inspirals}
\label{srcsec:emri}
MBHs in the centres of galaxies are typically surrounded by clusters of stars. Interactions between stars in these clusters can put objects onto orbits that pass very close to the central black hole, which leads to their gravitational capture by and eventual inspiral into the MBH. Compact objects (white dwarfs, neutron stars and stellar mass black holes) are sufficiently compact that they are not tidally disrupted but gradually inspiral via GW emission before finally plunging into and merging with the MBH. For MBHs in the appropriate mass range these extreme-mass-ratio inspirals (EMRIs) will generate millihertz GWs~\cite{EMRIrev}. LISA would have been able to observe these systems out to redshifts $z \apgt 1$, with an expected event rate of a few hundreds to several thousand per year~\cite{gair2009}. eLISA would be able to observe EMRIs out to redshift $z \sim 0.7$ with an expected event rate of several tens per year~\cite{eLISA,tgu}. In both cases, the event rate is dominated by inspirals of black holes, partially due to the enhancement of the intrinsic rate for these sources due to mass segregation and partially because they are detectable to greater distance. There are considerable uncertainties in these event rates due to the poorly understood physics of dense stellar clusters, but even a single EMRI observation could have a profound impact on our understanding of fundamental physics (see Section~\ref{sec:fundphys}).

\subsubsection{Ultra-compact binaries in the Milky Way}
\label{srcsec:wdb}
The majority of stars are formed in systems containing two or more stellar objects. Roughly half of the stellar binaries that form are sufficiently compact that they remain bound through the whole process of stellar evolution and evolve into compact systems containing white dwarfs, neutron stars or black holes. The shortest period ($\sim 1$hr) systems, known as ultra-compact binaries, generate millihertz gravitational waves~\cite{nelemansUCB}. eLISA would be able to resolve several thousand individual systems with S/N greater than 7 and measure the first time derivative of the frequency for many systems~\cite{nissanke2012}. LISA would be able to individually resolve as many as twenty thousand individual systems, measuring first derivatives of the orbital frequency for many and  second derivatives for a handful~\cite{nissanke2012}. At frequencies below about $2$mHz, there are so many ultra-compact binaries that the population creates an unresolvable stochastic foreground. This was expected to dominate over instrumental noise for LISA, but will most likely lie below the eLISA noise, although eLISA could possibly still detect it from its annual modulation~\cite{tgu}.

\subsubsection{Cosmological sources}
\label{srcsec:cos}
Processes occurring on high energy scales in the very early Universe can generate GWs at all frequencies. The typical frequency scale is determined by the horizon size at the epoch when the GWs were created and for the millihertz band this corresponds to the TeV energy scale. TeV GWs could be generated by phase transitions or by inflationary reheating in certain braneworld scenarios, allowing a millihertz GW detector to probe Higgs self-couplings, the presence of supersymmetry or conformal dynamics at TeV scales and the dynamics of compact extra dimensions~\cite{binetruy}.  eLISA will be able to place bounds on the energy density of relic TeV gravitational radiation at the level $\Omega_{\rm GW} \sim 10^{-5}$~\cite{eLISA,eLISAGWN,tgu}. For LISA this bound would be a factor of a few better. Phase transitions often create one-dimensional topological defects, cosmic strings, which can generate GWs from cusps (points on the string which are travelling at nearly the speed of light formed by the interaction of waves propagating along string loops in different directions) or kinks. Millihertz GW detectors offer the best prospects for detecting and constraining the properties of these cosmic string networks~\cite{tgu}.

\section{Science applications: astrophysics}
\label{sec:astro}
From observations of millihertz gravitational wave sources it will be possible to determine the systems parameters very precisely. This offers great potential to constrain the astrophysics of the sources. eLISA measurements of MBH mergers should be able to determine the masses of the two black holes to a fractional precision of $\sim10^{-3}$--$10^{-2}$, the spin of the primary/secondary to a fractional precision of $\sim 10^{-2}/10^{-1}$, the sky location of the source to approximately $100$ deg$^2$ and the luminosity distance to a few tens of percent~\cite{PEtask}. Measurements with LISA would be moderately (a factor of $2$--$3$) better for the intrinsic parameters, but considerably better (a factor of $10$) for the sky position and distance, thanks primarily to the addition of a second independent data stream due to the third arm. Extremely precise measurements of the parameters of EMRIs are also possible. For a source observed with a S/N of $30$, LISA observations would be expected to determine the masses of the two objects to a part in $10^4$, the central black hole spin magnitude to a few parts in $10^4$, the sky position to a few square degrees and the luminosity distance to $\sim10\%$~\cite{BCPE}. This parameter precision comes from being able to track the EMRI waveform phase over many hundreds of thousands of waveform cycles and so equally precise measurements are possible with eLISA, for a source at the same S/N. Typical S/N's for eLISA will be lower than for LISA, but by less than a factor of $2$ and since the expected parameter errors scale like the inverse of S/N, the order of magnitude expected for eLISA errors is comparable. 

Very precise measurements for individual sources will be interesting but do not encode much astrophysical information. Astrophysical results will come from looking at the statistics of the set of sources of a particular type that are observed. We will discuss some specific questions that millihertz gravitational wave observations will be able to address in the following sections.

\subsection{MBH constraints on growth of structure}
There is general consensus about the overall process of hierarchical structure formation. Galaxies were formed very early in the Universe (the most distant known galaxy is at a redshift $z \sim 10.7$~\cite{firstgal}) and from shortly after their formation many galaxies already contained massive black holes in their centres. The largest of these black holes are seen as QSOs at redshifts as high as $z\sim7.08$~\cite{firstqso}. Galaxies evolve through a sequence of mergers over cosmic time, while their black holes are expected to merge following mergers of the host and also grow through accretion. Several tight correlations exist between the properties of galaxies and the massive black holes they contain, in particular the $M$-sigma relation between the central black hole mass and the velocity dispersion of stars in the centre of the galaxy~\cite{msigma}. These correlations are indicative of the close co-evolution of a MBH and its host.

While this overall picture is widely accepted, there are some uncertainties, for instance in the initial mass distribution of the seed black holes from which the MBHs have grown, in the metallicity of the material from which the MBHs formed, in the efficiency of accretion and in the accretion geometry. Several models exist which can reproduce current observational constraints, but these constraints are mostly at the high-mass and low-redshift end of the black hole distribution, which is the regime accessible to current electromagnetic observations. The electromagnetically accessible regime is unlikely to change significantly over the next twenty years. Different models make different predictions for the high-redshift and low-mass regime of the black hole distribution, which is precisely the regime that will be explored by millihertz gravitational wave observations. GW observations with eLISA could therefore have a significant impact on our understanding of structure formation.

The constraints that LISA observations could place on models of MBH growth were explored in detail in~\cite{sgbv2011}. They considered two different seeding mechanisms --- light `POPIII' seeds, formed at high redshift with typical mass $\sim100M_\odot$, or heavy `quasistar' seeds, formed form direct collapse of dust at lower redshift and with mass $\sim10^5M_\odot$ --- two different metallicities for the dust from which the MBHs form --- $Z=0$ or a distribution across all $Z$ --- two different accretion models --- accretion always at the Eddington limit or using the more complex Merloni-Heinz prescription~\cite{MHacc} --- and two different accretion geometries --- `chaotic' accretion, in which small amounts of mass are added sequentially from random orientations, or `efficient' accretion, in which all mass is added in a single coherent accretion episode. The analysis considered LISA measurements of MBH mass and redshift only and considered two different detection criteria ($S/N$ detection threshold of $8$ or $20$) and two different detector configurations (1 data stream or 2). All possible model comparisons were considered and it was found that for most pairs of models three months of LISA observations would be sufficient to distinguish the two models with $95\%$ confidence. Models that differed only in the accretion efficiency could not be distinguished with a three month observation, although the predictions of such model pairs differ primarily in the predicted spins of the black holes, which were not included in the analysis. With only three months of observations it was also not possible to distinguish between the zero-metallicity/light-seed model and the metallicity-distribution/heavy-seed model. However, after one year of observation we would expect that all model pairs could be confidently distinguished.

The distinguishability of a subset of these models using eLISA observations has also been considered and is summarised in Table~\ref{tab:eLISAMBHmodel}. Using only mass and redshift measurements, eLISA can confidently distinguish the seed mass properties, but not the accretion efficiency. Once spin measurements are included, however, it should be possible to confidently distinguish all four models. 

\begin{table}
\begin{center}
\begin{tabular}{l|cccc|cccc}
&\multicolumn{4}{c|}{Without spins}&\multicolumn{4}{c}{With spins}\\
&LE&LC&HE&HC&LE&LC&HE&HC\\\hline
LE&$\times$&0.48&0.99&0.99&$\times$&0.96&0.99&0.99\\
LC&0.53&$\times$&1.00&1.00&0.13&$\times$&1.00&1.00\\
HE&0.01&0.01&$\times$&0.79&0.01&0.01&$\times$&0.97\\
HC&0.02&0.02&0.22&$\times$&0.02&0.02&0.06&$\times$\\
\end{tabular}
\end{center}
\caption{\label{tab:eLISAMBHmodel}Probability of choosing the row model over the column model with better than $95\%$ confidence when the row model is true (upper half table) or when the column model is true (lower half table). We consider models that differ in black hole seed prescriptions --- either light (L) or heavy (H) seeds --- and in the accretion efficiency --- either efficient accretion (E) or chaotic accretion (C). The results in the left-hand portion of the table labelled `without spins' are based on measurements of the black hole mass and redshift only, while those in the right-hand portion labelled `with spins' are based on measurements of the spin as well as mass and redshift.}
\end{table}

\begin{figure}
\sidecaption
\includegraphics[width=0.75\textwidth]{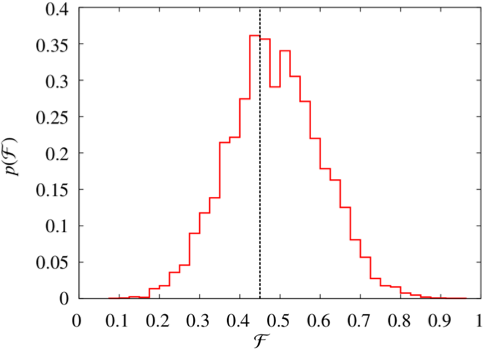}
%
%
\caption{Posterior distribution recovered by eLISA for the mixing fraction of a combination of the LE and HE models used in Table~\ref{tab:eLISAMBHmodel}. The injected model, a $0.45:0.55$ mix of LE:HE, is indicated by the vertical line. Figure reproduced from~\cite{eLISA}.}
\label{fig:eLISAmixing}       
\end{figure}

It is very unlikely that the Universe will be precisely described by one of these existing models, but instead all the processes will be taking place in some unknown proportion and the resulting evolution will represent some amalgam of possible scenarios. This was also explored in~\cite{sgbv2011}, in two different ways. In the first approach, mixed catalogues of mergers were constructed by linearly combining the catalogues from two or four different models in relative proportions $f_i \in [0,1]$, with $\sum f_i = 1$. This was done in two different ways --- either including the relative number of events predicted by the model when combining the two catalogues, or combining the  parameter distributions renormalised to one event and then marginalising over the number of observed events in the analysis. In each case, using both open and blind analyses, it was found that LISA observations would be able to derive posterior distributions on the mixing fractions that were consistent with the injected values and had typical widths of $\sim \pm 0.1$. eLISA will also be able to distinguish these mixed models, albeit with slightly less precision, $\sim \pm 0.15$ (see Figure~\ref{fig:eLISAmixing}). This linear mixing of catalogues is artificial since it does not include cross-mergers of black holes from the two different models. The second model mixing analysis in~\cite{sgbv2011} used numerical simulations to directly compute mergers for a hybrid model that included both light and heavy black hole seeds. Analysis of the hybrid models demonstrated that the linear-mixing model could successfully identify that the hybrid models comprised a mixture of light and heavy seed models and correctly indicated that the hybrid model with more efficient quasistar seeding had a greater fraction of heavy seeds.

More work is needed to fully understand the implications of eLISA observations for characterising the physics of structure formation. For instance, rather than merely determining the relative fractions of some distinct discrete models that are consistent with the observed data, it would be interesting to see how well eLISA could measure physical parameters that characterise the growth of structure, for instance the efficiency of quasistar formation, accretion efficiency etc. Nonetheless, it is already clear that eLISA will have a profound impact on our understanding of these processes, as it is extremely difficult to constrain these lower mass merging black holes using any other techniques.

\subsection{EMRI measurements of the MBH mass function}
EMRI observations will be primarily restricted to the low redshift Universe, $z \aplt 0.7$. The `cosmic high noon'~(as described by the CANDELS team~\cite{highnoon,highnoonB}) of peak star formation rate and maximal QSO activity occurred at redshifts $z\sim3$--$1.5$ and EMRI observations will therefore probe low mass black holes during the period of decline in AGN activity. Quiescent black holes in the relevant mass range are again poorly constrained observationally. Black hole masses in the regime $M < 10^7M_\odot$ can be inferred from observed galaxy luminosity functions using correlations between galaxy luminosity and stellar velocity dispersion and between velocity dispersion and central black hole mass~\cite{greeneho}. It is also possible to directly measure velocity dispersions for some galaxies using galaxy surveys, e.g., SDSS~\cite{sdss}. However, the luminosity function is incomplete at lower luminosities and the sensitivity limit of direct velocity dispersion measurements approximately corresponds to a black hole mass of $10^7M_\odot$. EMRI observations will therefore provide the first direct measurement of the low-mass end of the black hole mass function in the local Universe.

In~\cite{gtv} the ability of LISA EMRI observations to recover the parameters of a simple power-law black hole mass function, $dn/d\log M = A M^\alpha$, was considered. The analysis assumed that a threshold S/N of $30$ would be required for EMRI detection, used a circular-equatorial EMRI waveform model for fixed values of the central black hole spin of $a=0$ or $a=0.9$ and considered both a `pessimistic' (1 data channel, 2 year observation) and `optimistic' (2 data channels, 5 year observation) configuration of LISA. The precision with which LISA would be able to determine the parameters of the mass function was found to scale simply with one over the square root of the number of detected events, $N_{\rm obs}$. In the worst case (MBH spin $a=0$ and pessimistic detector assumptions) typical errors were $\Delta \ln A \approx 0.8 \sqrt{10/N_{\rm obs}}$ and $\Delta \alpha \approx 0.3 \sqrt{10/N_{\rm obs}}$, while for the best case (MBH spin $a=0.9$ and optimistic detector assumptions) the typical errors were $\Delta \ln A \approx 0.5 \sqrt{10/N_{\rm obs}}$, $\Delta \alpha \approx 0.2 \sqrt{10/N_{\rm obs}}$. The corresponding results for eLISA are similar to the pessimistic detector case, $\Delta \ln A \approx 1.1 \sqrt{10/N_{\rm obs}}$ and $\Delta \alpha \approx 0.35 \sqrt{10/N_{\rm obs}}$. Current electromagnetic constraints suggest that the black hole mass function is flat in log mass in the eLISA range, with an uncertainty in slope of $\pm 0.3$. Therefore, with just ten EMRI observations we will constrain the black hole mass function to a precision better than it is currently known. We expect eLISA to observe several tens of EMRI events per year, so it is likely that the derived constraint will be a considerable improvement on our current knowledge.

eLISA is actually sensitive to the product of the number density of black holes with the EMRI rate per black hole and the preceding analysis assumed that the latter took a known form and scaled with black hole mass to the power of $-0.15$, which is implied by numerical simulations. This is a source of uncertainty in the result, although it is hoped that over the next two decades this scaling will be constrained with sufficient precision that the degeneracy between the mass function and EMRI rate scaling can be broken. EMRI constraints on a redshift-dependent mass function of the form $dn/d\log M = A_0 (1+z)^{A_1} M^{\alpha_0-\alpha_1 z}$ were also considered in~\cite{gtv}, but the conclusion was that LISA would only be able to place very weak constraints on $A_1$ and $\alpha_1$. eLISA observations will be restricted to a smaller range of redshifts than those for LISA and so it is unlikely that eLISA EMRI observations will place interesting constraints on the evolution of the mass function.

\subsection{Other astrophysical applications}
Millihertz GW observations will also provide a rich variety of astrophysical information in other areas that have so far not been studied in detail. The analyses described above have considered the impact of MBH mergers or EMRIs alone, but the MBHs into which EMRIs are falling are the quiescent, low-redshift remnants of the same population that eLISA will observe merging at higher redshift. By using observed MBH merger and EMRI events in combination it will be possible to learn more about these systems. Combined observations might, for example, break the degeneracy between the EMRI rate and black hole mass function and provide insight into the evolution of the MBH mass function. In addition, MBH merger observations will provide constraints on the magnitudes and directions of the spins of the component black holes, which will offer indirect constraints on the processes that drive spin evolution, in particular accretion, and on the efficiency of mechanisms of spin alignment. EMRI observations will also provide spin measurements, which will provide insights into the spin distribution and accretion history of quiescent low-mass black holes in the local Universe. The characteristics of the orbits (eccentricity and inclination) of observed EMRIs encode information about the relative importance of different EMRI formation mechanisms --- EMRIs started through dynamical capture onto a highly eccentric orbit will have moderate eccentricity and random inclination in band; EMRIs started by the tidal splitting of a binary star or through tidal stripping of the envelope of a massive star will be circular but with non-zero inclination; and EMRIs that occur after the formation of stars in a disc around the central MBH will tend to be both circular and of zero inclination~\cite{EMRIrev}. EMRI measurements of compact object masses will reveal the properties of stellar populations in galactic central clusters and the efficiency of mass segregation in bringing heavier compact objects to the centre of the cluster. Observations of ultra-compact galactic binaries will provide a comprehensive survey of the population of these objects in the Milky Way, offering new constraints on stellar evolution. The binaries for which frequency derivatives are measured will provide information on the physics of mass transfer and of tidal interactions. Measuring the frequency derivative will also allow the distance to the source to be determined and we expect to have about $100$ such sources concentrated in the inner Galaxy, allowing a direct measurement of the distance to the Galactic centre to be made~\cite{tgu}. Finally, although eLISA will not directly observe coalescing compact binaries, the statistics of the observed population will significantly improve estimates of the binary merger rate.


\section{Science applications: cosmology}
\label{sec:cos}
As described in Section~\ref{srcsec:cos}, a millihertz GW detector could detect a stochastic background of radiation generated in the very early Universe, which would place strong constraints on TeV physics. We will not describe this further here, but refer the interested reader to~\cite{binetruy} for a comprehensive discussion of potential cosmological GW backgrounds in the millihertz frequency range. 
What we will briefly discuss is the use of millihertz GW sources as ``standard sirens'' to indirectly probe the expansion history of the Universe. The basic idea was first proposed by Bernard Schutz~\cite{schutzH0}. The observed strain of a GW source scales as $h \sim (1+z){\cal M}/D_L(z)$, where ${\cal M}$ is the chirp mass, $z$ the redshift and $D_L(z)$ the luminosity distance to the source. A GW detector can determine ${\cal M}$ very accurately by tracking the phase of the waveform, and so, if the mass-redshift degeneracy can be broken, the observed amplitude of the GW source can be used to determine $D_L(z)$ and hence probe the expansion history. The primary complication is to determine the redshift. This can be done if an electromagnetic counterpart to an event is observed. MBH mergers were thought to be promising candidates, in particular so-called `golden binaries' observed with high S/N at relatively low redshift~\cite{holzhughes}. However, no robust counterpart mechanism is currently known and weak lensing dominates the distance measurement error for sources at $z \apgt 1$. The weak lensing error can be mitigated to some extent by mass reconstruction, but not sufficiently to reach useful precision~\cite{hilbertWL}.

An alternative and more promising approach to GW cosmography is to combine measurements of multiple events statistically. This was the approach advocated by Schutz in his original paper~\cite{schutzH0} and it was first explored in the context of space-based detectors by McLeod and Hogan~\cite{McLeodHogan}. They considered measurements of the Hubble constant, $H_0$, using LISA observations of EMRIs. Given an observed LISA EMRI, a galaxy catalogue is used to identify the redshifts of all galaxies consistent with the LISA error box for that EMRI. For each galaxy, this redshift distribution is used to construct a distribution of distance estimates and hence cosmological parameters. These distributions are then combined for all the events observed. Although employing galaxy catalogues, the method does not rely on the fact that the true host galaxy is in the catalogue, but will work as long as the distribution of EMRI host galaxies is similar to the distribution of observed galaxies. McLeod and Hogan found that LISA observations of $\sim 20$ EMRI events at redshifts $z < 0.5$ would be sufficient to determine the Hubble constant to $\sim 1\%$. This calculation has not been repeated for eLISA. We expect eLISA to detect the required $\sim 20$ events at redshift $z<0.5$, but eLISA's typical distance errors will be larger. Assuming, conservatively, that the eLISA error would be a factor of $2$ times larger, we would expect to determine $H_0$ to $\sim 2\%$ with $20$ EMRI events or to $\sim 1\%$ with $80$ events. We expect to observe a few tens of EMRI events per year with eLISA, so this number would only be reached with an extended mission. In~\cite{PBScos} a similar analysis was carried out for the constraints that LISA observations of MBHs could place on cosmological parameters. They found that LISA would be able to place constraints on the equation of state of dark energy that are a factor of $2$--$8$ better than current constraints from electromagnetic observations. The analysis has not been repeated for eLISA, but eLISA determination of MBH merger distances is somewhat poorer so it is likely that eLISA constraints will be no better than those based on current electromagnetic data.

As eLISA is not scheduled for launch until $2034$ it is likely that any constraints on cosmological parameters derived from eLISA observations will not be competitive with the best constraints obtained from other methods by that time. However, these gravitational wave constraints will be completely independent of all other measurements, which will provide important confirmation of existing results and could, in principle, reveal the presence of previously unappreciated systematics in the electromagnetic results.

\section{Science applications: fundamental physics}
\label{sec:fundphys}
Gravitational wave observations offer a unique way to probe the strong-field, dynamical regime of gravity that has so far not been constrained experimentally. All GW observations can be used to test the theory of general relativity (GR), but the strongest constraints will come from millihertz sources. MBH mergers are the loudest GW events that we expect to observe, while EMRIs should be detectable for hundreds of thousands of waveform cycles over the last several years of inspiral. A one cycle phase difference over the observation should be detectable in principle, allowing very small fractional deviations from the model, $\aplt 10^{-5}$, to be identified. The potential impact of a space-based gravitational wave detector on our understanding of fundamental physics has been discussed extensively in the literature. We will briefly summarise some of the key ideas in the following, but refer the reader to~\cite{TestGRLRR} for a thorough review of this topic.

\subsection{Tests of gravitational physics}
\label{sec:testgravphys}
GW observations will provide tests of several different aspects of gravitational physics.

{\bf GW polarisation}: In GR, there are only two possible polarisation states for gravitational waves --- the transverse tensor `plus' and `cross' modes (TT) --- but up to four additional polarisation modes  --- scalar-transverse (ST), scalar-longitudinal (SL) and two vector-longitudinal (VL) modes --- may exist in alternative metric theories of gravity~\cite{willtestGR}. At high frequencies, LISA would be ten times more sensitive to SL and VL  modes than to ST and TT modes~\cite{tinto2010}, although whether these would be detected in practice depends on the relative amplitude at which the different modes are generated by a given source in a particular alternative theory.

{\bf GW propagation}: In GR, GWs are predicted to travel at the speed of light. GW observations can measure the speed of GWs and hence constrain the effective `graviton mass' either by comparing the GW arrival time to that of an electromagnetic counterpart or by detecting the dispersion of gravitational wave chirps (different frequency components travel at different speeds). Bounds on the Compton wavelength of the graviton of $\lambda_g =h/m_g\apgt \mbox{a few} \times 10^{16}$km should be possible using the population of observed eLISA MBH merger events~\cite{bgsMG}.

{\bf Inspiral rate}: If energy is lost from a system to other forms of radiation in addition to the TT modes of GR this will manifest itself as a difference in the observed rate of inspiral of a system. In scalar-tensor gravity dipole radiation is generated in addition to the TT modes. The best GW bounds on this dipole radiation will come from observations of neutron star-MBH inspirals observed with space-based detectors. The derivable bound on the coupling constant of Brans-Dicke gravity scales as~\cite{willyunesBD}
\begin{equation}
\omega_{\rm BD} > 2 \times 10^4 \left( \frac{{\cal S}}{0.3} \right)^2 \left( \frac{100}{\Delta\Phi_D} \right) \left( \frac{T}{1 {\rm yr}} \right)^{\frac{7}{8}} \left( \frac{10^4 M_\odot}{M_\bullet} \right)^{\frac{3}{4}} 
\end{equation}
where ${\cal S}$ is the `sensitivity', the difference between the self-gravitational binding energy per unit rest mass of the neutron star and the MBH, $\Delta \Phi_D$ is the dipole contribution to the GW phase over the observation, $T$ is the observation time and $M_\bullet$ is the MBH mass. This is comparable to the current bound obtained from the Cassini spacecraft, but including the effects of spin-orbit coupling and eccentricity, ignored in the preceding equation, can reduce the bound by as much as a factor of ten~\cite{yagitanakaBD}.

{\bf Phase evolution}: A general gravitational wave can be written in the frequency domain in the form $\tilde{h}(f) = {\cal A}(f) \exp[i \Psi(f)]$. GR makes definite predictions for the coefficients in an expansion of the phase function in powers of frequency. A natural thing to do in order to provide sensitivity to un-modelled departures from GR is to try to measure these coefficients directly, or include extra terms in the expansion. Two different approaches to this problem have been explored. The first is to consider the coefficients, $\{\psi_k\}$, of the expansion of the phase in powers of $f^{(k-5)/3}$ to be free parameters to be measured from the data~\cite{arunppnA}. For an MBH merger source at a distance of $3$Gpc, LISA would be able to measure $\psi_0$ to $\sim0.1\%$ and $\psi_2$ and $\psi_3$ to $\sim 10\%$, but no other parameters. Alternatively, if $\psi_0$ and $\psi_2$ are used to determine the masses of the two objects in the binary and then the other coefficients are just checked for consistency with GR, all of the parameters can be constrained to $\sim \mbox{a few }\%$~\cite{arunppnB}. The second approach, the so-called `parameterised post-Einsteinian' formalism (ppE)~\cite{ppE}, in its simplest form, involves multiplying the GR waveform by a factor $(1 + \alpha (\pi {\cal M}f)^a) \exp[i\beta(\pi{\cal M}f)^b]$ and attempting to constrain the parameters $\alpha, a, \beta, b$ that represent the leading order departures in the amplitude and phase from GR. All currently known alternative theories of gravity can be represented in this ppE formalism. It was shown in~\cite{cornishppE} that for $b \apgt-1.75$, LISA would be able to place more stringent constraints on the amplitude of the deviation parameter $\beta$ than is possible with either binary-pulsar observations or the post-Newtonian coefficient modification approach described above. The ppE formalism has now been generalised to alternative polarisation states, higher harmonics and eccentricity (see description and references in~\cite{TestGRLRR}).

\subsection{Tests of the nature and structure of black holes}
\label{sec:testBHnature}
In GR, with certain additional assumptions made on physical grounds (the space-time is stationary, vacuum, asymptotically flat and contains an event horizon, but no closed time-like curves exterior to the horizon), the unique end state of gravitational collapse is a Kerr black hole~\cite{hawkingellis}. The Kerr metric depends on just two free parameters --- the mass of the black hole, $M$, and its (dimensionless) angular momentum, $a$. All higher multipole moments of the Kerr metric are determined by these two quantities, $M_l + iS_l = M(i a)^l$, so this uniqueness result is sometimes referred to as the `no-hair' theorem. EMRIs are expected to be observed on eccentric and inclined orbits for hundreds of thousands of GW cycles  while the inspiralling object is in the strong field region of the space-time. The small body thus explores the whole strong-field regime and the emitted GWs encode a detailed map of the space-time structure which can be used to make precise tests of the consistency of the space-time with the Kerr metric. This idea was first discussed in depth in~\cite{ryan95}, who showed that the multipole moments of the space-time were encoded in GW observables --- the precession frequencies of the orbit and the number of waveform cycles generated at a given frequency. Different multipole moments enter at different orders in the expansion of the precession frequencies as functions of the orbital frequency and so by observing the evolution of the frequencies over an inspiral a map of the space-time can be constructed. Many aspects of space-time mapping have been explored over the subsequent two decades and we summarise some of the key results in the following. Most of these results have been computed in the context of observations with LISA. However, the key thing that they rely on is the ability to track EMRI phase over the inspiral, which will be equally possible with eLISA. The conclusions are therefore unlikely to be changed by modifications of the mission design.

{\bf The nature of the central object}: In~\cite{ryan97} it was shown that LISA could measure the first three multipole moments of the space-time with reasonable precision, which is enough for a test of the no-hair property. However, as more multipole moments were added in the model, the expected precision with which the lower multipole moments could be determined was degraded. The multipole moment expansion is an inefficient way to carry out space-time mapping, as an infinite number of multipoles are required to characterise the expected Kerr solution. Subsequent studies, starting with~\cite{CH2004}, focussed on `bumpy black holes', space times which are Kerr plus a small perturbation. Perturbative modifications of Schwarzschild~\cite{CH2004} and Kerr~\cite{GB2006} have been considered as well as exact solutions to the field equations of GR that violate one of the physicality assumptions of the uniqueness theorem~\cite{GLM2008} and the linear spin black hole space time predicted in dynamical Chern Simons modified gravity (a parity-violating modification to GR inspired by string theory)~\cite{SY2009}. Fisher matrix analyses of the precision with which LISA observations would be able to determine the size of the deviations from GR have been carried out for quadrupole-deformed space-times~\cite{BCbumpy} and the Chern Simons deformation~\cite{CGS2012}. These studies provided a consistent conclusion that LISA EMRI observations would be able to measure deviations from the Kerr metric at the level of $\sim 0.1\%$, while simultaneously measuring the mass and spin to $\sim 0.01\%$. LISA could also place a bound on the parameter characterising the Chern-Simons deviation at the level of $\xi^{\frac{1}{4}} < 10^4$km, four orders of magnitude better than the current Solar System bound~\cite{CGS2012}.

Differences in the central object from a Kerr black hole are also encoded in qualitative features of the emitted GWs. If the object does not have a horizon, e.g., if it were a compact boson star, GW emission could continue after the object `plunges' into the horizon~\cite{KGK2005}. Energy is lost into the horizon through tidal coupling and so information about this tidal coupling can be inferred from comparing the observed inspiral rate to the flux of energy carried to infinity in the gravitational waves. LISA would be able to distinguish an $\sim O(1)$ proportional change in the tidal coupling efficiency~\cite{LiLove2008}. Finally, differences in the nature of the horizon also manifest themselves in the quasi-normal mode (QNM) structure, e.g., `gravastars' in which the horizon is replaced by a thin shell~\cite{pani2009}. These QNMs can be excited by an EMRI, leading to resonances in the observed inspiral evolution. The excitement of QNMs following a black hole merger can also be used to constrain deviations from GR, which will be discussed further below.

{\bf The effect of matter}: Departures in our observations from the model predictions could also come about from astrophysical perturbations. Various possibilities have been considered. In~\cite{barausseA} the gravitational influence of a torus of material on an EMRI was investigated and it was found to be significant. However, making small adjustments to the mass and spin of the space-time (which are unknown and have to be measured from the data) made the signals indistinguishable. The effect of hydrodynamic drag from a disc around an MBH was considered in~\cite{barausseB} and estimated to be detectable, but only for very compact discs around low-mass MBHs. The signature of the drag is a decrease in orbital inclination, which is opposite to the evolution under GW emission. EMRIs inspiraling within a disc were considered in~\cite{Yunesdisc} and it was found that a $\sim 1$ radian dephasing relative to a vacuum EMRI could accumulate over an observation in some circumstances. The effect of a distant massive perturber (another MBH within the same galaxy in a hierarchical triple) was considered in~\cite{YMT2011} and it was found that a perturber within $\sim 0.1$pc could leave a measurable imprint. A second EMRI occurring in the same system could also leave an imprint in $\sim1\%$ of EMRIs by causing nominally chaotic evolution of the orbit~\cite{butterfly}. Finally, the presence of exotic material, e.g., axion~\cite{axioncloud} or boson~\cite{bosoncloud} clouds, could leave measurable imprints on the signal. While each of these processes could leave an imprint, they should not occur in much more than a few per cent of observed EMRIs. Moreover, the qualitative signature of an astrophysical perturbation should be different from an intrinsic deviation in the central object, as the former should become weaker as the inspiral progresses, while the latter should become stronger. Further study may be required, but it is likely that astrophysical perturbations will not adversely impact tests of fundamental physics. For a comprehensive discussion of environmental impacts on GW tests of relativity we refer the interested reader to~\cite{BCPenviron}.

{\bf Strong field dynamics}: Departures from GR could lead to large qualitative differences in the behaviour of the orbits and emitted waves in the strong-field regime. Kerr has a complete set of integrals of the motion, which means the equations of motion are separable and the orbits are tri-periodic. A generic deviation from Kerr will break this property in general. Chaotic orbits have been found in several non-Kerr space times, for instance~\cite{GLM2008}. Observing the signs of chaos would be a clear smoking-gun for a deviation from Kerr, but it is not obvious how to detect chaos in practice, since LISA data analysis will use EMRI templates based on Kerr, which will be for tri-periodic and predictable orbits. However, if an inspiral passed from a regime where it was Kerr-like and tri-periodic and then transitioned into a chaotic regime, this would be observable as an EMRI `disappearing' from the data prematurely~\cite{GLM2008}. Another qualitative signature of a departure from Kerr would be the behaviour of the EMRI as it passes through a resonance on which the EMRI frequencies become commensurate. Resonances persist generically when integrable systems are perturbed~\cite{persistres}, so the resonance would last much longer than expected.

{\bf Generic deviations from GR}: In the spirit of the ppE formalism described in Section~\ref{sec:testgravphys}, it is interesting to consider the imprint of generic deviations from GR in the GWs generated during EMRIs. In~\cite{GY2011} EMRI waveforms were constructed for a set of modified gravity space-times described in~\cite{VYS2011} that were built to be close to Kerr but maintain a complete set of integrals of the motion. These were thought to be the most relevant types of deviation as separable inspirals will be most easily detected using a Kerr EMRI analysis for the reasons discussed in the preceding section. This family includes the linear-spin metric of Chern Simons modified gravity mentioned earlier~\cite{SY2009}. These metrics are characterised by the scaling of the perturbation with radius. LISA should be able to constrain the size of the deviation to a precision of $\sim 10^{-6}$ for an $r^{-2}$ departure from GR, with the precision of the bound degrading by a factor of $10$ for each additional order of $r$ in the perturbation. 

{\bf Quasi-normal ring down radiation}: A perturbed black hole settles down to a stationary state by emitting gravitational radiation that is a superposition of quasi-normal modes. Each QNM is a damped sinusoid, characterised by a frequency and damping time which, due to the no-hair property, depend on the mass and spin of the black hole only. A measurement of two QNMs will therefore provide a consistency check that could reveal departures from the Kerr solution. EMRIs do not excite significant QNMs due to the large mass-ratio, but QNMs excited by MBH mergers can be used for no-hair tests. In~\cite{bertiQNM} it was concluded that it should be possible for LISA to resolve one QNM and either the frequency or damping time of a second QNM, provided that the second mode radiates at least $\sim10^{-2}$ of the total ringdown energy. Numerical relativity simulations suggest that this will be the case~\cite{bertiQNM2}. Using these simulations it was estimated that a ringdown S/N of $\sim 30$ would be needed to carry out a no-hair test~\cite{bertiQNM3} and therefore that a $1\%$ departure from the QNMs of GR could be identified for a $10^8M_\odot$ black hole at $50$Gpc and a $10\%$ departure for a $10^6M_\odot$ black hole at $6$Gpc~\cite{gossanQNM}. A full review of studies of no-hair tests using LISA observations of QNMs can be found in Section 6.3 of~\cite{TestGRLRR}.

\section{Summary}
\label{sec:discuss}
The millihertz GW band contains a rich variety of sources which have tremendous potential for science. Observations in this band could transform our understanding of astrophysics, cosmology and fundamental physics. They will provide constraints on the properties of MBHs and their evolution over cosmic time and on stellar populations in galactic centres and throughout the Milky Way. Millihertz GW sources can be used to probe the expansion history of the Universe in a manner that is completely independent of all current constraints from electromagnetic observations. Finally, these observations will push tests of GR into a regime of strong-fields and dynamical gravity that has never been probed. We will tests all aspects of gravitational physics and probe the nature and structure of black hole space times to an unprecedented precision. {\it The Gravitational Universe}~\cite{tgu} has been selected by ESA as the science theme for the L3 large satellite mission to launch in $\sim2034$. This mission will open up the millihertz gravitational wave window for the first time and enable us to realise all of this outstanding scientific potential.

\begin{acknowledgement}
JG's work is supported by the Royal Society.
\end{acknowledgement}

\end{document}